\documentclass[pre,showpacs,amsmath,amsfonts,endfloats]{revtex4}

\usepackage{bm}

\newtheorem{theorem}{Theorem}
\newtheorem{lemma}{Lemma}

\newtheorem{algorithm}{Algorithm}

\begin{document}

\title{Geometry and Topology of Escape II: Homotopic Lobe Dynamics}

\author{K.~A.~Mitchell}
\email{kevinm@physics.wm.edu}
\author{J.~P.~Handley} 

\affiliation{Department of Physics, College of William and Mary,
Williamsburg, Virginia, 23187-8795}

\author{S.~K.~Knudson} 
\affiliation{Department of Chemistry, College of William and Mary,
Williamsburg, Virginia, 23187-8795}

\author{J.~B.~Delos} 
\email{jbdelo@wm.edu}

\affiliation{Department of Physics, College of William and Mary,
Williamsburg, Virginia, 23187-8795}

\date{\today}

\begin{abstract}
We continue our study of the fractal structure of escape-time plots
for chaotic maps.  In the preceding paper, we showed that the
escape-time plot contains regular sequences of successive escape
segments, called epistrophes, which converge geometrically upon each
endpoint of every escape segment.  In the present paper, we use
topological techniques to: (1) show that there exists a minimal
required set of escape segments within the escape-time plot; (2)
develop an algorithm which computes this minimal set; (3) show that
the minimal set eventually displays a recursive structure governed by
an ``Epistrophe Start Rule'': a new epistrophe is spawned $\Delta =
D+1$ iterates after the segment to which it converges, where $D$ is
the minimum delay time of the complex.
\end{abstract} 

\maketitle

Topological methods and symbolic dynamics have long been valuable
tools for describing orbits of dynamical systems.  For example, if a
particle in the plane scatters from three fixed disks, labeled A,B,
and C, its orbit can be characterized by a sequence of symbols, such
as ...ABA*BCBCA..., giving the sequence of collisions with the disks.
The asterisk gives the location of the particle at the present time;
as time goes by the asterisk takes one step to the right.  In this
paper, we describe a new kind of symbolic dynamics, in which the
symbol sequence describes the structure of a curve in the plane.  The
relevant curve is not the trajectory of a particle, but rather an
ensemble of initial points in phase space -- the line of initial
conditions.  This line winds around ``holes'' in the plane in a manner
described by the symbol sequence.  The dynamical map applied to the
line induces a map on the symbol sequence, which is more complicated
than a simple shift.  We use this symbolic dynamics to derive
properties of the epistrophes introduced in the preceding paper.  In
particular, we use it to obtain a ``minimal set of escape segments''
and an ``epistrophe start rule.''

\section{Introduction}

As in the preceding paper \cite{Mitchell03} (Paper I), we study maps
of the phase plane, which have an unstable fixed point (an
\textsf{X}-point) and an associated homoclinic tangle of stable and
unstable manifolds (Fig.~\ref{fPSP}).  The stable and unstable
manifolds intersect transversely, bounding a region of phase space
(called the ``complex'') from which a trajectory may or may not
escape.  We consider an initial distribution of points along a curve
passing through the complex (the line of initial conditions).  The
escape-time plot is the number of iterates of the map required for a
point to escape the complex, plotted as a function along the line of
initial conditions (Fig.~\ref{fETP}).  For chaotic systems, such
escape-time plots have a complicated set of singularities and
structure at all levels of resolution
\cite{Noid86,Petit86,Eckhardt86,Eckhardt87,Jung87,Gaspard89,Gaspard98,Ruckerl94a,Ruckerl94b,Lipp95,Jung99,Butikofer00,Tiyapan93,Tiyapan94,Tiyapan95}.
These fractal escape-time plots play a central role in a variety of
classical decay and scattering problems; we have been particularly
motivated by the ionization of atoms, especially hydrogen in parallel
electric and magnetic fields.

The escape-time plot of a discrete map is divided into ``escape
segments'', intervals over which the escape time is constant.  In
Paper I, we proved that there exist certain important sequences of
consecutive escape segments, which we called epistrophes, at all
levels of resolution.  The epistrophes are characterized by the
Epistrophe Theorem, whose core results are: (1) each endpoint of an
escape segment spawns a new epistrophe which converges upon it; (2) in
the limit $n_i \rightarrow \infty$, every epistrophe converges
geometrically, with rate equal to the Liapunov factor $\alpha$ of the
\textsf{X}-point; (3) the asymptotic tails of any two epistrophes
differ by a simple scaling.

The focus of Paper I was the asymptotic behavior of epistrophes.  In
the present paper, we address how epistrophes begin.  We use the
topological structure of the homoclinic tangle and the line of initial
conditions to show that there is a certain minimal set of escape
segments.  For this minimal set, we prove the Epistrophe Start Rule
(Theorem \ref{t1}), which says that after a sufficiently large time,
each epistrophe begins some number of iterates $\Delta$ after the
segment that spawned it (i.e. the segment upon which the epistrophe
converges.)  The number $\Delta$ is the same for all epistrophes of a
given map and is dependent on the topological structure of the tangle;
explicitly, $\Delta = D + 1$, where $D$ is the minimum delay time of
the complex, that is, the minimum number of iterates a scattering
trajectory spends inside the complex.  The bulk of our effort is
devoted to developing an algebraic algorithm for constructing the
minimal set of escape segments for a general line of initial
conditions.  This algorithm allows us to compute the initial structure
of the escape-time plot for iterates before the Epistrophe Start Rule
sets in.  The algorithm is then used to prove the Epistrophe Start
Rule itself.

A critical aspect of this paper is our use of homotopy theory.  We
develop the necessary formalism in Sect.~\ref{sHLD}, and prior
knowledge of homotopy theory is not required.  Homotopy theory
provides an algebraic framework for describing the topological
structure of curves in the phase plane.  As we shall explain, the
phase plane has a set of ``holes'' into which the line of initial
conditions cannot pass.  A symbol sequence can be used to describe how
the line circumvents these holes.  As the dynamics maps the line
forward, there is an induced dynamics on the symbol sequence,
representing a new kind of symbolic dynamics which we call ``homotopic
lobe dynamics''.  From the symbol sequence, one can readily read off
the structure of the minimal set of escape segments.  Lines with
different symbol sequences may have different minimal sets; however,
at long enough times, these minimal sets always obey the Epistrophe
Start Rule.

Some escape segments, such as that marked with an asterisk in
Fig.~\ref{fETP}, are not within the minimal set guaranteed by the
topology.  These segments are ``surprises'' which, within the present
topological analysis, we cannot predict.  Since they break the regular
structure and since they often have no obvious connection with any of
the epistrophes, we call them ``strophes'' as in Sect. IIIB of Paper
I.  Strophes interfere with the self-similar structure of the fractal
and typically do not go away in the asymptotic limit, resulting in
what we called ``epistrophic self-similarity'' in Sect.~IIIC of Paper
I.  Despite the presence of these strophes, the minimal set often
seems to accurately describe the early and intermediate time structure
of the escape-time plot.

Patterns similar to our Epistrophe Start Rule have been seen in other
work.  In the numerical study of Tiyapan and Jaff\'e\cite{Tiyapan93},
epistrophes and the Epistrophe Start Rule are evident in the structure
of the initial angle-final action plot (analogous to the escape-time
plot).  Similarly, Jung and coworkers
\cite{Ruckerl94a,Ruckerl94b,Lipp95} used symbolic dynamics to
construct a tree-diagram that gives a comparable description of a
scattering system.  In each case, the authors consider a specific line
of initial conditions that is far outside of the complex and is
topologically simple.  Easton \cite{Easton86}, followed by Rom-Kedar
and others \cite{Rom-Kedar90,Rom-Kedar94,Litvak-Hinenzon95}, showed
that recursive patterns also apply to homoclinic intersections between
the stable and unstable manifolds.  Thus, it may not be surprising
that comparable patterns should apply to the intersections between the
stable manifold and an arbitrary line of initial conditions, at least
at sufficiently large iterate.  But at what iterate does this pattern
set in, and what is the minimal set before it sets in?  Algorithm
\ref{a1} answers both these questions, as well as giving a simple
proof of the Epistrophe Start Rule.  An important observation is that
the escape-time plot depends both on the topology of the tangle and on
the topology of the line of initial conditions.

The paper is organized as follows.  Section \ref{sNumerics} motivates
our study by presenting numerical computations on a particular
saddle-center map with a chosen line of initial conditions.  Section
\ref{sHLD} is the theoretical heart of the paper, in which we formally
develop homotopic lobe dynamics.  Section \ref{sAlgorithm} contains
Algorithm \ref{a1} for computing the minimal set of escape segments.
Section \ref{sESR} contains Theorem \ref{t1}, which includes the
Epistrophe Start Rule.  In Sect.~\ref{sExamples} we apply our
techniques to the escape-time plots for two representative lines of
initial conditions.  Conclusions are in Sect.~\ref{sConclusions}.
Appendices \ref{sAlgProof} and \ref{sT1proof} contain the proofs of
Algorithm \ref{a1} and Theorem \ref{t1} respectively.
Table~\ref{table1} summarizes the notation in this article.

\section{Numerical Data for an Example System}

\label{sNumerics}

As an example we study the map $\mathcal{M}$ defined by Eqs.~(A1) --
(A3) of Paper I using parameter values $\tau = 1.5$, $f = 0.25$, $m =
0.57$.  Figure \ref{fPSP} shows a phase space portrait for this map,
along with the line of initial conditions $\mathcal{L}_0$ considered
here.  The same map is plotted in Fig.~1, Paper I, but with a
different line of initial conditions.

We review the basic picture of phase space transport described in
Paper I and Refs.~\cite{MacKay84,Wiggins92,Rom-Kedar90,Rom-Kedar94}.
The map $\mathcal{M}$ has an unstable fixed point (\textsf{X}-point)
denoted $\mathbf{z}_\textsf{X}$ and having Liapunov factor $\alpha>1$,
which is the larger of the two eigenvalues of $\mathcal{M}$ linearized
about $\mathbf{z}_\textsf{X}$.  The \textsf{X}-point has an associated
homoclinic tangle consisting of the branch $\mathcal{S}$ of the stable
manifold and the branch $\mathcal{U}$ of the unstable manifold
(Fig.~\ref{fPSP}).  The {\em complex} is the region of phase space
bounded on the north by the segment of $\mathcal{S}$ connecting the
homoclinic intersection $\mathbf{P}_0$ to $\mathbf{z}_\textsf{X}$ and
bounded on the south by the segment of $\mathcal{U}$ connecting
$\mathbf{P}_0$ to $\mathbf{z}_\textsf{X}$.  The forward and backward
iterates $\mathbf{P}_n = \mathcal{M}^n(\mathbf{P}_0)$ are homoclinic
intersections with the same sense as $\mathbf{P}_0$.  The homoclinic
intersection $\mathbf{Q}_0$ and its iterates $\mathbf{Q}_n =
\mathcal{M}^n(\mathbf{Q}_0)$ have the opposite sense.

The {\em escape zone} $E_0$ is the lobe bounded by the segments of
$\mathcal{S}$ and $\mathcal{U}$ joining $\mathbf{P}_0$ to
$\mathbf{Q}_0$.  It maps forward to the lobes $E_n$, $n \ge 0$, which
all lie outside the complex, and backward to the lobes $E_{-n}$,
$n>0$, which all intersect the complex.  Similarly, the {\em capture
zone} $C_0$ is the lobe bounded by the segments of $\mathcal{S}$ and
$\mathcal{U}$ between $\mathbf{Q}_{-1}$ and $\mathbf{P}_0$.  It maps
forward to the lobes $C_n$, $n > 0$, which all intersect the complex,
and backward to the lobes $C_{-n}$, $n \ge 0$, which all lie outside
the complex.  Under one iterate of the map the escape zone $E_{-1}$
maps from inside to outside the complex and the capture zone $C_0$
maps from outside to inside the complex; the lobes $E_{-1}$ and $C_0$
together form what is called a {\em turnstile}
\cite{MacKay84,Wiggins92}.  It is important to emphasize that all
points which escape in $n$ iterates lie in the escape zone $E_{-n}$.

In the escape-time plot shown in Fig.~\ref{fETP}, the number of
iterates $n_i$ to escape is plotted as a function along the line of
initial conditions $\mathcal{L}_0$.  Figure \ref{fETP} is analogous to
Fig.~2 of Paper I, but for a different choice of
$\mathcal{L}_0$.  For a given $n_i$, the set of escaping points is
partitioned into open intervals called {\em escape segments}; an
escape segment is one connected component of $E_{-n_i} \cap
\mathcal{L}_0$.  For example, the first three escape segments at $n_i
= 3,4,5$ in Fig.~\ref{fETP} correspond to the three intersections of
$\mathcal{L}_0$ with the lobes $E_{-n}$, $n = 3,4,5$, shown in
Fig.~\ref{fPSP}.

The Epistrophe Theorem of Paper I says that the escape-time plot
contains sequences of escape segments, called epistrophes.  Several
epistrophes are denoted by arrows in Fig.~\ref{fETP}.  The first
epistrophe starts at $n_i = 3$ and converges monotonically upward,
containing one escape segment for each $n_i \ge 3$.  A second
epistrophe begins at $n_i = 9$ and converges downward upon the
endpoint of the $n_i = 3$ segment.  We say that the $n_i = 3$ segment
``spawns'' this epistrophe.  Two more epistrophes are spawned at $n_i
= 10$ and converge upon the two endpoints of the $n_i = 4$ segment.
Similarly, the $n_i = 5$ segment spawns two more epistrophes beginning
at $n_i = 11$.

The data in Fig.~\ref{fETP} suggest the following Epistrophe Start
Rule: each endpoint of an escape segment at $n$ iterates spawns an
epistrophe which begins at $n + \Delta$ iterates, where in this case
$\Delta = 6$.  This recursive rule is formulated precisely by Theorem
\ref{t1} in Sect.~\ref{sESR}.  In general, $\Delta = D+1$, where $D$
describes the global topology of the tangle (Sect.~\ref{sGroupoid}).
The fact that $\Delta = 6$ in Fig.~\ref{fETP} is a consequence of the
fact that $E_{-3}$ intersects $C_3$ (and no earlier $C_n$, $n<3$) in
Fig.~\ref{fPSP}.

On the left of Fig.~\ref{fETP} are plotted the winding numbers $n_w$
of the escaping trajectories, i.e., the number of times a given
trajectory winds around the central stable zone as it escapes to
infinity.  Notice that all segments of the epistrophe beginning at
$n_i = 3$ have winding number $n_w = 1$.  Similarly, all segments of
the epistrophes spawned by the $n_i = 3, 4, 5$ segments have winding
number $n_w = 2$.  The data in Fig.~\ref{fETP} thus suggests that all
segments of an epistrophe have the same winding number and that this
number is one greater than the winding number of the segment which
spawned the epistrophe.  This rule will be precisely formulated and
proved in a separate publication.

\section{Homotopic Lobe Dynamics}

\label{sHLD}

We introduce a new kind of symbolic dynamics, where the symbol
sequences refer to paths in the plane (rather than trajectories of the
map.)  This symbolic dynamics allows us to identify and describe a
minimal set of escape segments along an arbitrary line of initial
conditions.  The theory of homotopy is central to our
development~\cite{Nakahara90,Massey77,Fomenko86}.  Homotopy theory
allows us to ignore the detailed positions of the stable and unstable
manifolds and concentrate instead on their global topological
structure.  Homotopy theory also provides a natural algebraic
framework for describing this global structure.

We consider a ``saddle-center map'' $\mathcal{M}$, which has a simple
homoclinic tangle, as seen in Fig.~\ref{fPSP} and described precisely
by Assumptions 1 -- 5 in Paper I \cite{footnote10}.

\subsection{The Homotopy Groupoid}

\label{sGroupoid}

We define the {\em active region} $A$ to be the union of the complex
with all of its forward and backward iterates.  By construction, it is
an invariant region of the phase plane.  The boundary of $A$, denoted
$\partial A$, contains alternating segments of $\mathcal{S}$ and
$\mathcal{U}$ (the outer boundaries of capture and escape zones) as
well as the \textsf{X}-point \cite{footnote1}.  The boundary $\partial
A$ has a well-defined orientation determined by the orientations of
$\mathcal{S}$ and $\mathcal{U}$.

Let $D>0$ be the smallest integer such that $C_D$ intersects $E_0$.
Considering all scattering trajectories which begin outside the
complex, enter the complex, and eventually exit, $D$ is the smallest
possible number of iterates spent inside the complex.  For this
reason, we call $D$ the {\em minimum delay time} of the complex or
simply the {\em delay time}.  The delay time is equivalently defined
by the first pre-iterate $E_{-D}$ of $E_0$ which intersects $C_0$.  In
Fig.~\ref{fPSP}, the delay time $D$ is equal to 5.  (The delay time
$D$ agrees with Easton's signature $\ell$
\cite{Easton86,Rom-Kedar90,Rom-Kedar94}.)

For the case $D=1$, shown in Fig.~\ref{fSOSD1}, $E_{-1}$ intersects
$C_1$, forming an open region $H_{-1} = E_{-1} \cap C_1$, which we
view as a {\em hole} in the active region $A$.  Mapping this hole
backwards and forwards gives an infinite set of holes $H_n$.  More
generally, for arbitrary $D$ we define the holes $H_n = E_n \cap
C_{n+\Delta}$, $-\infty < n < \infty$, where $\Delta = D+1$.  (See
Fig.~\ref{fSOSD3} for the case $D=3$.)  The set $A^* = A\setminus
\cup_n H_n$ is the active region minus all the holes $H_n$.  The $D$
holes $H_{-D}, ..., H_{-1}$ are inside the complex; all other holes
are outside.

The homoclinic intersections $\mathbf{P}_n$ and $\mathbf{Q}_n$, $-\infty < n <
\infty$, form a subset $\alpha$ of the boundary $\partial A$.  Two
paths (or directed curves) having the same initial and final points
$\mathbf{s}_0, \mathbf{s}_1 \in \alpha$ are said to be {\em homotopic} if one
can be continuously distorted into the other without passing through a
hole $H_n$ and without moving their endpoints \cite{footnote2}.  The
concept of homotopy defines equivalence classes of paths; the
path-class, or {\em homotopy class}, $a$ is the set of all paths
homotopic to an arbitrary path $\mathcal{A} \in a$.  That is, two
paths belong to the same homotopy class if they can be distorted one
into the other without changing the endpoints or passing through any
hole; likewise, two homotopy classes $a$ and $b$ are equal if a path
$\mathcal{A}$ in class $a$ can be distorted into a path $\mathcal{B}$
in class $b$.

We are particularly interested in the following paths.  For each $n$,
we define $\mathcal{S}_n$ to be the path along the
$\mathcal{S}$-boundary of capture zone $C_n$, joining $\mathbf{Q}_{n-1}$
to $\mathbf{P}_n$, and we define $\mathcal{U}_n$ to be the path along the
$\mathcal{U}$-boundary of escape zone $E_n$, joining $\mathbf{P}_n$ to
$\mathbf{Q}_n$, as shown in Fig.~\ref{fSOSD1}.  These paths lie in the
boundary $\partial A$ of the active region.  Similarly, for each $n$,
we define $\mathcal{E}_n$ to be the path along the
$\mathcal{S}$-boundary of $E_{n}$ and $\mathcal{C}_n$ to be the path
along the $\mathcal{U}$-boundary of $C_{n}$.  These paths bound the
lobes $E_n$ and $C_n$ in the interior of $A$.  Since each path
$\mathcal{E}_n$, $\mathcal{C}_n$, $\mathcal{U}_n$, and $\mathcal{S}_n$
has endpoints in $\alpha$ and does not pass through any of the holes
$H_n$, each belongs to a well-defined homotopy class.  These classes
are distinct, since none of the curves can be distorted into any
other, and we denote them by $e_n$, $c_n$, $u_n$, and $s_n$,
respectively.  These homotopy classes encode global topological
information about the structure of the tangle.

Let $\Pi(A^*, \alpha)$ be the collection of all homotopy classes of
paths in $A^*$ having endpoints in $\alpha$.  For a path-class $a_1
\in \Pi(A^*, \alpha)$ joining $\mathbf{s}_0$ to $\mathbf{s}_1$ and a
path-class $a_2 \in \Pi(A^*, \alpha)$ joining $\mathbf{s}_1$ to
$\mathbf{s}_2$, their product $a_1a_2$ joins $\mathbf{s}_0$ to $\mathbf{s}_2$ and
is constructed by first traversing a representative path
$\mathcal{A}_1 \in a_1$ followed by a representative $\mathcal{A}_2
\in a_2$.  The homotopy class of a constant path, i.e. one which
remains at a given point $\mathbf{s} \in \alpha$ for all times, is denoted
$1$ (with the endpoint $\mathbf{s} \in \alpha$ being understood from
context); for all $a \in \Pi(A^*, \alpha)$, $1 a = a 1 = a$.  For a
class $a \in \Pi(A^*, \alpha)$, its inverse $a^{-1}$ contains a
representative path from $a$, but traversed backwards; clearly, $a
a^{-1} = 1$.  The set $\Pi(A^*, \alpha)$ thus has most of the
structure of a group (multiplication, identity, and inverse) except in
one respect: the product $a_1a_2$ is not defined between arbitrary
elements $a_1$ and $a_2$ but only between elements such that the final
point of $a_1$ equals the initial point of $a_2$.  A set with such a
restricted product is called a groupoid \cite{MacKenzie87}, and $\Pi(A^*, \alpha)$ is
called the {\em fundamental groupoid of path-classes in $A^*$ having
base points in $\alpha$}.

The dynamical map $\mathcal{M}$, acting on points in the plane,
induces a map on the path-classes, which forms a kind of symbolic
dynamics on the symbols $e_n$, $c_n$, $u_n$, and $s_n$.  When
$\mathcal{M}$ acts on these elements, it simply shifts their indices,
\begin{subequations}
\begin{eqnarray}
\mathcal{M}(e_n) & = & e_{n+1}, \label{r11}\\
\mathcal{M}(c_n) & = & c_{n+1}, \label{r8} \\
\mathcal{M}(u_n) & = & u_{n+1}, \label{r6} \\
\mathcal{M}(s_n) & = & s_{n+1}. \label{r12}
\end{eqnarray}
\label{r1}
\end{subequations}

\subsection{The Minimal Set of Escape Segments}

\label{sMinimalSet}

We now turn our attention to the line of initial conditions, which we
assume is given by a path $\mathcal{L}_0$ that (1) has endpoints
$\mathbf{\lambda}_{i}, \mathbf{\lambda}_{f} \in \partial A$, (2) does not
self-intersect, and (3) does not intersect any hole $H_n$ or the
\textsf{X}-point $\mathbf{z}_{\textsf{X}}$ \cite{footnote7}.  For the
homotopy analysis, we must shift the endpoints of $\mathcal{L}_0$ so
that they lie in the set $\alpha$.  For example, we shift the initial
point by first traversing a path $\mathcal{K}_i$ before traversing
$\mathcal{L}_0$; $\mathcal{K}_i$ begins at one of the two points in
$\alpha$ on either side of $\mathbf{\lambda}_{i}$, runs along $\partial
A$, and finally terminates at $\mathbf{\lambda}_{i}$.  By thus shifting
both endpoints, we assign to $\mathcal{L}_0$ a well-defined homotopy
class $\ell_0 \in \Pi(A^*, \alpha)$ \cite{footnote3}.

The intersection of $\mathcal{L}_0$ with escape zone $E_{-n}$ is the
set of points that escape on the $n$th iterate, and any connected
component of this set is called an {\em escape segment}; sometimes we
will use the term {\em $e_{-n}$-segment} to emphasize an intersection
with $E_{-n}$.  (The index $n$ may, in fact, be either positive or
negative.)  In this article, we answer the following two questions
regarding a minimal set of escape segments.

\noindent \textbf{Question 1} {\em What is the minimum number of
intersections possible between a representative path $\mathcal{L}_0'
\in \ell_0$ and a representative path $\mathcal{E}_{-n}' \in e_{-n}$?}

The minimum number of $e_{-n}$-segments is half the minimum number of
intersections.

\noindent \textbf{Question 2} {\em Let $\mathcal{L}_0' \in \ell_0$,
$\mathcal{E}_{-n_1}' \in e_{-n_1}$, $\mathcal{E}_{-n_2}' \in e_{-n_2}$
($n_1 \ne n_2$) be paths which minimize all possible pairwise- and
self-intersections.  In particular, $\mathcal{E}_{-n_1}'$ and
$\mathcal{E}_{-n_2}'$ do not intersect each other or themselves, and
$\mathcal{L}_0'$ has the minimum number of escape segments at both
$n_1$ and $n_2$ iterates.  What are the positions of the escape
segments at $n_1$ iterates relative to those at $n_2$ iterates?
\cite{footnote8} }

In answering these questions we allow ourselves to distort
$\mathcal{L}_0$ and $\mathcal{E}_{-n}$ into paths $\mathcal{L}_0'$ and
$\mathcal{E}_{-n}'$ to minimize the number of intersections.  Thus we
are constructing a ``distorted escape zone'' $E_{-n}'$ whose
intersection with $\mathcal{L}_0'$ is a set of ``distorted escape
segments''.  Henceforth, we omit the descriptor ``distorted'' and
leave it understood.

The answer to the above two questions will be obtained from the
algebraic algorithm in Sect.~\ref{sAlgorithm}, which will lead in turn
to a proof of the Epistrophe Start Rule in Sect.~\ref{sESR}.

\subsection{The Untangled Basis  of Path-Classes}

\label{sUB}

By a {\em basis} of a groupoid we mean a minimal set of elements that
generate the entire groupoid.  To construct a basis of the fundamental
groupoid $\Pi(A^*, \alpha)$, we first include the path-classes $(...,
s_{-1}, s_0, s_{1}, ... \; ; \; ..., u_{-1}, u_0, u_{1}, ...  )$ along
the boundary of the active region $\partial A$.  We then select
path-classes $( ..., c_{-1}, c_0 \; ; \; c_1, ..., c_D \; ; \; e_0,
e_1, ...)$ that encircle the holes in $A^*$, so that the complete
basis is 
\begin{equation}
( ..., c_{-1}, c_0\; ; \; c_1, ..., c_D\; ; \; e_0, e_1,
... \; ; \; ..., s_{-1}, s_0, s_{1}, ... \; ; \; ..., u_{-1}, u_0,
u_{1}, ...  ), 
\label{r7}
\end{equation}
shown schematically in Fig.~\ref{fBC}.  The elements
$c_1, ..., c_D$ are special in that they are the only basis elements
which must enter the interior of the complex, encircling the $D$ holes
$H_{-D}, ..., H_{-1}$.

The representative paths $\mathcal{C}_n$, $\mathcal{E}_n$,
$\mathcal{S}_n$, and $\mathcal{U}_n$ for this basis satisfy (see
Fig.~\ref{fBC}): (1) no path in the basis intersects itself or any
other path in the basis (except perhaps at the end points); (2) each
representative $\mathcal{E}_n$ and $\mathcal{C}_n$ in the basis
encircles exactly one hole, and each hole is encircled exactly once.
Furthermore, (3) all homotopy classes of relevance to us, specifically
$\ell_0$, $c_n$, and $e_n$, $-\infty < n < \infty$, have a unique {\em
finite} reduced expansion in the basis \cite{footnote9}.  (A reduced
expansion is a sequence of elements in which any two adjacent factors
$a$ and $a^{-1}$ have been canceled.)  Because of these properties and
the simple picture shown in Fig.~\ref{fBC}, we call this basis the
``untangled basis''.

\subsection{Symbolic Dynamics of Path-Classes}

\label{sAlgorithm}

Now we develop the symbolic dynamics that will describe the minimal
set of escape segments.  First, however, we must assign a direction to
each escape segment.  Recall that the two endpoints of an
$e_n$-segment ($-\infty < n < \infty$) are intersection points between
a path $\mathcal{E}_n' \in e_n$ and a path $\mathcal{L}' \in \ell$.
Using the orientation of $\mathcal{E}_n'$, one of these endpoints
occurs first.  We define the direction of an escape segment to point
along $\mathcal{L}'$ from the {\em second} endpoint to the {\em first}
endpoint.  (See Fig.~\ref{fdir}.)  Recall that $\mathcal{L}'$ has an
independent direction defined by its own parameterization.  An escape
segment is said to ``point forward'' if its direction is the same as
$\mathcal{L}'$ and to ``point backward'' otherwise.  A point on
$\mathcal{L}'$ is said to lie on the ``positive'' side of an escape
segment if the segment points toward it and on the ``negative'' side
otherwise.

We need the forward iterates of all untangled basis elements expressed
in terms of the untangled basis.  For most elements, this is given by
Eqs.~(\ref{r1}).  Only one additional equation is needed
\begin{equation}
\mathcal{M}( c_D ) 
= c_{D+1} 
= F^{-1} u_0^{-1} e_0 F s_{D+1}, 
\label{r15} 
\end{equation}
where $F$ is an abbreviation for the path-class
\begin{equation}
F = c_1 e_1 c_2 e_2 ... c_D e_D.
\label{r16}
\end{equation}
Notice that the right-hand side of Eq.~(\ref{r15}) [after substituting in
Eq.~(\ref{r16})] is expressed entirely in terms of the untangled basis
(\ref{r7}).  Equation (\ref{r15}) is proved by first observing
\begin{equation}
e_{-1} 
= u_{-1} (c_0 e_0 F e_{D}^{-1}) s_D^{-1} c_D (e_D F^{-1} e_0^{-1} c_0^{-1}), 
\label{r33}
\end{equation}
which, though rather lengthy, can be directly verified from a figure
such as \ref{fSOSD1} or \ref{fSOSD3}; one simply concatenates the
basis paths as shown on the right and then distorts the resulting path
into $\mathcal{E}_{-1} \in e_{-1}$.  By applying $\mathcal{M}$ to both
sides of Eq.~(\ref{r33}) and solving for $c_{D+1}$, one obtains
Eq.~(\ref{r15}).  It is convenient to explicitly compute the forward
iterate of $F$ from Eq.~(\ref{r15})
\begin{equation}
\mathcal{M}(F) = e_1^{-1}c_1^{-1}u_0^{-1}e_0 F s_{D+1}e_{D+1}.
\label{r9}
\end{equation}

For the purpose of computing the minimal set of escape segments, the
$e_n$ basis elements ($n \ge 0$) and the $s_n$ basis elements (all
$n$) can simply be omitted from any expression that contains them,
resulting in significant computational simplification; for example,
Eqs.~(\ref{r15}), (\ref{r16}), and (\ref{r9}) above become
Eqs.~(\ref{r2}) -- (\ref{r3}) below.  This is explained more fully in
Appendix~\ref{sAlgProof}.  We can now state the algorithm for
constructing the minimal set of escape segments up to a given
iterate $N$.

\begin{algorithm}

\label{a1}

Let $\mathcal{L}_0$ be the line of initial conditions and $\ell_0 \in
\Pi(A^*, \alpha)$ its homotopy class.

\noindent 1) Expand $\ell_0$ in the untangled basis, omitting any
$e_n$-factors for $n \ge 0$ and all $s_n$-factors for $-\infty < n <
\infty$.

\noindent 2) Compute $\ell_N$ by iterating $\ell_0$ forward $N$ times
using Eqs.~(\ref{r8}), (\ref{r6}), and
\begin{equation}
\mathcal{M}( c_D ) = F^{-1} u_0^{-1} F,
\label{r2}
\end{equation}
where 
\begin{equation}
F = c_1 c_2 ... c_D.
\label{r10}
\end{equation}
For convenience, one may also use the following formula, which
explicitly maps $F$ forward,
\begin{equation}
\mathcal{M}(F) = c_1^{-1} u_0^{-1} F.
\label{r3}
\end{equation}
Carry out any cancellations of factors using $a a^{-1} = 1$, so that
$\ell_N$ is expressed as a reduced expansion in the untangled basis.

Then,

\noindent A) Each $u_{n}$ or $u_{n}^{-1}$ factor ($n \ge 0$) in the
expansion of $\ell_N$ corresponds to a segment that escapes in $n_i =
N - n$ iterates and that points backwards or forwards, respectively.

\noindent B) The relative positions of the $u_{n}$-factors in the
expansion of $\ell_N$ are the same as the relative positions of their
corresponding escape segments along $\mathcal{L}_0$.

\end{algorithm}
This algorithm is justified in Appendix~\ref{sAlgProof}.

\subsection{Examples}

We apply Algorithm~\ref{a1} to compute the minimal set of escape
segments (up to $n_i = 3$) for the simple example $D=1$, $\ell_0 = c_D
= c_1$.  Carrying out Step 2, the first three iterates of $\ell_0$ are
computed to be
\begin{subequations}
\begin{eqnarray}
\ell_0 & = & c_1, \\
\ell_1 & = & c_1^{-1} \underline{u_0^{-1}} c_1, \\
\ell_2 & = & 
c_1^{-1} \underline{u_0} c_1 \underline{u_1^{-1}} c_1^{-1} \underline{u_0^{-1}} c_1, \label{r4} \\
\ell_3 & = & 
c_1^{-1} \underline{u_0} c_1 \underline{u_1} c_1^{-1} \underline{u_0^{-1}} c_1 \underline{u_2^{-1}} 
c_1^{-1} \underline{u_0} c_1 \underline{u_1^{-1}} c_1^{-1} \underline{u_0^{-1}} c_1,
\label{r34}
\end{eqnarray}
\end{subequations}
where the $u_n$-factors have been underlined for greater visibility.
We now consider the consequences of results A and B in the algorithm.
Examining $\ell_1$, it contains a single factor $u_0^{-1}$, which
yields a single forward pointing escape segment at $n_i=1$, as shown
in Fig.~\ref{fETPQ}a.  Iterating forward to $\ell_2$, this $n_i=1$
escape segment corresponds to the factor $u_1^{-1}$ in Eq.~(\ref{r4});
on either side of this factor are factors $u_0$ and $u_0^{-1}$,
corresponding respectively to backward and forward pointing segments
that escape at $n_i=2$.  Iterating once more, $\ell_3$ has four
$u_0$-factors (either $u_0$ or $u_0^{-1}$) corresponding to four
escape segments at $n_i=3$ and with relative positions and directions
shown in Fig.~\ref{fETPQ}a.

Considering now an arbitrary $D$, $\ell_0 = c_D$ propagates forward as
\begin{subequations}
\begin{eqnarray}
\ell_0 & = & c_D, \\
\ell_1 & = & c_{D+1} = F^{-1} \underline{u_0^{-1}} F, \\
\ell_2 & = & c_{D+2} = (F^{-1} \underline{u_0} c_1) \underline{u_1^{-1}} (c_1^{-1} \underline{u_0^{-1}} F), \\
\ell_3 & = & c_{D+3} = (F^{-1} \underline{u_0} c_1 \underline{u_1} c_2) \underline{u_2^{-1}} 
(c_2^{-1} \underline{u_1^{-1}} c_1^{-1} \underline{u_0^{-1}} F), 
\label{r22} \\
\ell_4 & = & c_{D+4} = (F^{-1} \underline{u_0} c_1 \underline{u_1} c_2 \underline{u_2} c_3) \underline{u_3^{-1}} 
(c_3^{-1} \underline{u_2^{-1}} c_2^{-1} \underline{u_1^{-1}} c_1^{-1} \underline{u_0^{-1}} F), \\
& \vdots & \nonumber \\
\ell_{n} & = & c_{D+n} = (F^{-1} \underline{u_0} c_1 \underline{u_1} c_2 ... \underline{u_{n-2}} c_{n-1}) 
\underline{u_{n-1}^{-1}} 
(c_{n-1}^{-1} \underline{u_{n-2}^{-1}} ... c_2^{-1} \underline{u_1^{-1}} c_1^{-1} \underline{u_0^{-1}} F), \\
& \vdots & \nonumber \\
\ell_{D+1} & = & c_{2D+1} = (F^{-1} \underline{u_0} c_1 \underline{u_1} c_2 ... \underline{u_{D-1}} c_D) \underline{u_D^{-1}} 
(c_D^{-1} \underline{u_{D-1}^{-1}} ... c_2^{-1} \underline{u_1^{-1}} c_1^{-1} \underline{u_0^{-1}} F), \\
\ell_{D+2} & = & c_{2D+2} = (F^{-1} \underline{u_0} c_1 \underline{u_1} c_2 ... \underline{u_{D-1}} c_D) 
(\underline{u_D} F^{-1} \underline{u_0^{-1}} F \underline{u_{D+1}^{-1}} F^{-1} \underline{u_0} F \underline{u_D^{-1}}) \nonumber \\
&& \;\;\;\;\;\;\;\;\;\;\; \;\; (c_D^{-1} \underline{u_{D-1}^{-1}} ... c_2^{-1} \underline{u_1^{-1}} c_1^{-1} \underline{u_0^{-1}} F). 
\label{r23}
\end{eqnarray}
\label{r19}
\end{subequations}
The minimal set of escape segments for $\ell_0 = c_D$, as constructed
from results A and B in the algorithm, is shown schematically in
Fig.~\ref{fETPQ}b.  The set contains a left- and a right-converging
epistrophe, with two additional segments at $n_i = D+2$.  These two
segments are the beginnings of two new epistrophes spawned $\Delta =
D+1$ iterates after the first segment.  This spawning behavior is also
visible in Fig.~\ref{fETPQ}a for $\Delta = D+1 = 2$.  In the next
section we show that all lines of initial conditions have a minimal
set that eventually displays such spawning behavior.

\subsection{The Epistrophe Start Rule}

\label{sESR}

After a certain number of iterates, the minimal set for any
$\mathcal{L}_0$ has a simple recursive structure described by the
following theorem, which is proved in Appendix \ref{sT1proof}.

\begin{theorem}

\label{t1}

Let $\mathcal{M}$ be a ``saddle-center map'' satisfying Assumptions 1
-- 5 of Paper I \cite{footnote10} and having an arbitrary minimum
delay time $D \ge 1$.  Let $\mathcal{L}_0$ be the line of initial
conditions.  There exists some iterate $N_0 >0$ such that the minimal
set of escape segments at all $N \ge N_0$ iterates can be constructed
from the following two recursive rules:

\vskip 12pt

\noindent (i) {\em Epistrophe Continuation Rule:} Every segment (in
the minimal set) that escapes at $N-1$ iterates has on its immediate
positive side a segment that escapes at $N$ iterates and which has
the same direction.

\noindent (ii) {\em Epistrophe Start Rule:} Every segment that
escapes at $N-\Delta$ iterates ($\Delta = D + 1$) spawns immediately
on both of its sides a segment that escapes at $N$ iterates and which
points toward the spawning segment.

\vskip 12pt
\noindent 
Explicitly, $N_0 = \max \{ -n_c+1, -n_u, 0 \} + D + 2$, where $n_c$
and $n_u$ are respectively the lowest indices of the $c_n$- and
$u_n$-factors in the expansion of the path-class $\ell_0$ of
$\mathcal{L}_0$ in the untangled basis.
\end{theorem}

To say that an $e_{n_1}$-segment lies ``on the immediate
positive/negative side of'' an $e_{n_2}$-segment means that in the
minimal set there is no earlier $e_{n_3}$-segment, $n_3 \le \max\{
n_1, n_2 \}$, between the two.  Notice that new epistrophes are
spawned by the Epistrophe Start Rule; the Epistrophe Continuation Rule
simply propagates those epistrophes started earlier.  Notice also that
segments of an epistrophe point in the direction of convergence of the
epistrophe.  The early structure of the minimal set (before $N_0$) can
be computed using the algorithm in Sect.~\ref{sAlgorithm}.  Thus, the
algorithm gives the early behavior of the minimal set, and the simpler
recursive rules give the subsequent behavior.

\section{Examples}

\label{sExamples}

Using the map $\mathcal{M}$ discussed in Sect.~\ref{sNumerics} and
illustrated by Fig.~\ref{fPSP}, we consider the escape-time plots for
two different lines of initial conditions.

\subsection{Line 1}

\label{sExample1}

We consider the line of initial conditions $\mathcal{L}_0$ in
Fig.~\ref{fPSP}.  First we determine the homotopy class of
$\mathcal{L}_0$.  Since neither endpoint of $\mathcal{L}_0$ is in
$\alpha$, we must shift each endpoint as described in
Sect.~\ref{sMinimalSet}.  Since the initial (southernmost) endpoint is
on the curve $\mathcal{U}_{-3}$ (the southern boundary of $E_{-3}$),
we can shift it either east to $\mathbf{P}_{-3}$ or west to $\mathbf{Q}_{-3}$;
we choose $\mathbf{Q}_{-3}$ since this will guarantee that the beginning
of $\mathcal{L}_0$ still intersects $E_{-3}$.  Since the final
(northernmost) endpoint is on the curve $\mathcal{S}_3$ (the northern
boundary of $C_3$), it does not matter whether we shift it east to
$\mathbf{P}_3$ or west to $\mathbf{Q}_2$; we choose $\mathbf{Q}_2$.  Following
Step 1 in the algorithm, we scrutinize Fig.~\ref{fPSP} to determine
that the homotopy class $\ell_0 \in \Pi(A^{*}, \alpha)$ of the
adjusted curve is $\ell_0 = c_{-2} u_{-2} c_{-1} u_{-1} c_0 e_0 F
e_5^{-1} s_5^{-1} e_4^{-1} s_4^{-1} e_3^{-1} s_3^{-1}$.  After
omitting $e_n$- and $s_n$-factors, this simplifies to
\begin{equation}
\ell_0 = c_{-2} u_{-2} c_{-1} u_{-1} c_0 F. 
\label{r61}
\end{equation}
Following Step 2, we map $\ell_0$ forward using Eqs.~(\ref{r8}),
(\ref{r6}), (\ref{r2}), and (\ref{r3}) with $D = 5$,
\begin{subequations}
\begin{eqnarray}
\ell_1 & = & c_{-1} u_{-1} c_{ 0} F, \\
\ell_2 & = & c_{ 0} F, \\
\ell_3 & = & \underline{u_0^{-1}} F, \\
\ell_4 & = & u_1^{-1} c_1^{-1} \underline{u_0^{-1}} F, \\
\ell_5 & = & u_2^{-1} c_2^{-1} u_1^{-1} c_1^{-1} \underline{u_0^{-1}} F, \\
\ell_6 & = & u_3^{-1} c_3^{-1} u_2^{-1} c_2^{-1} 
             u_1^{-1} c_1^{-1} \underline{u_0^{-1}} F, \\
\ell_7 & = & u_4^{-1} c_4^{-1} u_3^{-1} c_3^{-1} 
             u_2^{-1} c_2^{-1} u_1^{-1} c_1^{-1} 
             \underline{u_0^{-1}} F, \\
\ell_8 & = & u_5^{-1} c_5^{-1} u_4^{-1} c_4^{-1} 
             u_3^{-1} c_3^{-1} u_2^{-1} c_2^{-1} 
             u_1^{-1} c_1^{-1} \underline{u_0^{-1}} F, \\
\ell_9 & = & u_6^{-1} F^{-1} \underline{u_0} F  
             u_5^{-1} c_5^{-1} 
             u_4^{-1} c_4^{-1} u_3^{-1} c_3^{-1} 
             u_2^{-1} c_2^{-1} u_1^{-1} c_1^{-1} 
             \underline{u_0^{-1}} F.
\end{eqnarray}
\label{r60}
\end{subequations}
%
%
%
For greater visibility, we have underlined each $u_0$-factor.  Mapping
forward once more, we find
\begin{equation}
\begin{array}{ccccccccccccccccccccccccc}
\ell_{10} = 
 & u_7^{-1} & F^{-1} & u_0 & c_1 & 
 u_1 & c_1^{-1} & u_0^{-1} &
 F & u_6^{-1} & F^{-1} & u_0 & F & 
 u_5^{-1} & c_5^{-1} \times \\
n_i & \overrightarrow{3} && \overleftarrow{10} && \overleftarrow{9} && \overrightarrow{10} && \overrightarrow{4} && \overleftarrow{10} && \overrightarrow{5} &&  \\
\\
& u_4^{-1} & c_4^{-1} &
 u_3^{-1} & c_3^{-1} & u_2^{-1} & c_2^{-1} & 
 u_1^{-1} & c_1^{-1} & u_0^{-1} & F. \\
 & \overrightarrow{6} && \overrightarrow{7} && \overrightarrow{8} && \overrightarrow{9} && \overrightarrow{10}
\end{array}
\label{r50}
\end{equation}
Below each $u_n$-factor in Eq.~(\ref{r50}), we have recorded the
number of iterates to escape; the arrow indicates whether the segment
is forward- or backward-pointing.  The results of Eqs.~(\ref{r60}) and
(\ref{r50}) are shown qualitatively in Fig.~\ref{fSETP}a; they should
be compared with the calculation in Fig.~\ref{fETP}.  We examine these
results in detail.

(1) As stated in Algorithm \ref{a1}, each $u_n$ or $u_n^{-1}$ factor
    in $\ell_N$ corresponds to a segment of $\mathcal{L}_0$ that
    escapes in $n_i = N-n$ iterates.  Equation (\ref{r50}) gives the
    minimal set of escape segments up to $n_i = 10$.
(2) After a certain iterate $N_0$, we can determine the minimal set
    using the Epistrophe Continuation Rule and Epistrophe Start Rule
    in Theorem \ref{t1}.  Explicitly, $N_0 = \max \{ -n_c +1, -n_u, 0
    \} + D + 2$; examining Eq.~(\ref{r61}) we see $n_c = n_u = -2$,
    and since $D = 5$, $N_0 = 10$.  So, for all iterates $n_i \ge 10$,
    Algorithm \ref{a1} and Theorem \ref{t1} give identical results.
(3) Direct computation (Fig.~\ref{fETP}) indicates that up to $n_i =
    14$, there are no additional escape segments outside the minimal
    set.  The first segment in the computation which is not in the
    minimal set is indicated by an asterisk in Fig.~\ref{fETP} at $n_i
    = 15$; it is an example of what we call a strophe.
(4) No epistrophe converges upon the lower endpoint of the $n_i = 3$
    segment, either in the minimal set (Fig.~\ref{fSETP}a) or the
    numerical data (Fig.~\ref{fETP}), because this point is an
    intersection between $\mathcal{L}_0$ and the {\em unstable}
    manifold.

\subsection{Line 2}

\label{sExample2}

We consider the line of initial conditions $\mathcal{L}_0$ in Fig.~1
of Paper I.  In order to define the homotopy class of this line, it
must first be adjusted.  From Fig.~1, Paper I, we see that
$\mathcal{L}_0$ intersects the holes $H_{-1} = E_{-1} \cap C_5$ and
$H_{-5} = E_{-5} \cap C_1$.  We adjust $\mathcal{L}_0$ within each of
these holes so that it runs along the boundary of the hole, on either
the east or west side, and not through the hole itself.  For the
northern hole $H_{-5}$, we adjust $\mathcal{L}_0$ to run along the
eastern boundary, so that it still passes through $E_{-5}$.  For the
southern hole $H_{-1}$, we adjust $\mathcal{L}_0$ to run along the
western boundary.  As in Sect.~\ref{sExample1}, the endpoints of
$\mathcal{L}_0$ must also be adjusted, so that they lie in $\alpha$.
We shift the southern endpoint to $\mathbf{P}_{-1}$ and the northern
endpoint to $\mathbf{P}_1$.

The homotopy class $\ell_0 \in \Pi(A^{*}, \alpha)$ of the adjusted
curve is $\ell_0 = u_{-1} c_0 e_0 c_1$, which simplifies to
\begin{equation}
\ell_0 = u_{-1} c_0 c_1.
\label{r63}
\end{equation}
Then $\ell_0$ maps forward as
\begin{subequations}
\begin{eqnarray}
\ell_1 & = & \underline{u_{ 0}} c_1 c_2, \\
\ell_2 & = & u_{ 1} c_2 c_3, \\
\ell_3 & = & u_{ 2} c_3 c_4, \\
\ell_4 & = & u_{ 3} c_4 c_5, \\
\ell_5 & = & u_{ 4} c_5 F^{-1} \underline{u_0^{-1}} F, \\
\ell_6 & = & u_{ 5} F^{-1} c_1 u_1^{-1} c_1^{-1} \underline{u_0^{-1}} F, \\
\ell_7 & = & u_{ 6} F^{-1} \underline{u_0} c_1 c_2 
             u_2^{-1} c_2^{-1} u_1^{-1} 
             c_1^{-1} \underline{u_0^{-1}} F, \\
\ell_8 & = & u_{ 7} F^{-1} \underline{u_0} c_1 u_1 
             c_2 c_3 u_3^{-1} c_3^{-1} u_2^{-1} 
             c_2^{-1} u_1^{-1} c_1^{-1} \underline{u_0^{-1}} F, \\
\ell_9 & = & u_{ 8} F^{-1} \underline{u_0} c_1 u_1 c_2 u_2 
             c_3 c_4 {u_4^{-1}} c_4^{-1} {u_3^{-1}} 
             c_3^{-1} {u_2^{-1}} c_2^{-1} {u_1^{-1}} c_1^{-1} 
             \underline{u_0^{-1}} F, \\
\ell_{10}& = & {u_{ 9}} F^{-1} \underline{u_0} c_1 {u_1} c_2 
             {u_2} c_3 {u_3} 
             c_4 c_5  {u_5^{-1}} c_5^{-1} {u_4^{-1}} 
             c_4^{-1} {u_3^{-1}} c_3^{-1} {u_2^{-1}} c_2^{-1} 
             {u_1^{-1}} c_1^{-1} \underline{u_0^{-1}} F, 
\end{eqnarray}
\end{subequations}
and
\begin{equation}
\begin{array}{ccccccccccccccccccccccccc}
\ell_{11}& = & {u_{10}} & F^{-1} & {u_0} & c_1 & {u_1} & c_2 & 
             {u_2} & c_3 & {u_3} & c_4 & {u_4} &
             c_5 F^{-1} & {u_0^{-1}} & F & 
             {u_6^{-1}} \times \\
n_i && \overleftarrow{1} && \overleftarrow{11} && \overleftarrow{10} && \overleftarrow{9} && \overleftarrow{8} && \overleftarrow{7} && \overrightarrow{11} && \overrightarrow{5} \\
\\
&&           F^{-1} & {u_0} & F & {u_5^{-1}} & 
             c_5^{-1} & {u_4^{-1}} & c_4^{-1} & {u_3^{-1}} & 
             c_3^{-1} & {u_2^{-1}} & c_2^{-1} & {u_1^{-1}} & c_1^{-1} & 
             {u_0^{-1}} & F. \\
& && \overleftarrow{11} && \overrightarrow{6} && \overrightarrow{7} && \overrightarrow{8} && \overrightarrow{9} && \overrightarrow{10} && \overrightarrow{11} 
\end{array}
\label{r54}
\end{equation}
The data from Eq.~(\ref{r54}) are summarized in Fig.~\ref{fSETP}b.
This should be compared with the numerical calculation in Fig.~2 of
Paper I.

Equation (\ref{r54}) gives the minimal set of escape segments up
    to $n_i = 11$.  In this case, examining Eq.~(\ref{r63}), $n_c = 0$
    and $n_u = -1$, yielding $N_0 = \max \{ 1,1,0 \} +5 +2 = 8$.
    Therefore for $n_i \ge 8$, the minimal set can be generated from
    Theorem \ref{t1} rather than the algorithm.
The first numerically computed segment which is not in the minimal set
    (a strophe) does not occur until $n_i = 16$; it is indicated by an
    asterisk in Fig.~2 of Paper I.
As above, no epistrophe converges upon the lower endpoint of the $n_i
    = 1$ segment because it is an intersection between $\mathcal{L}_0$
    and the unstable manifold.

\section{Conclusions}

\label{sConclusions}

The results of the present paper combine with the results of Paper I
\cite{Mitchell03} to create a detailed picture of escape-time plots.
On the one hand, the present study predicts the existence of a minimal
set of escape segments (Algorithm \ref{a1}).  After some number of
iterates, this set has a simple recursive pattern (Theorem \ref{t1})
described by: (1) at each iterate, add new segments that perpetuate
all earlier epistrophes; (2) at $\Delta = D+1$ iterates after a given
segment, spawn two new epistrophes on either side of this segment.
These results say nothing about the lengths of segments or the
separation between segments, and in particular say nothing about
convergence properties of epistrophes.  On the other hand, the results
of Paper I do address such issues, and we find that epistrophes
converge geometrically upon the endpoints of the segments that spawn
them and furthermore that all epistrophes differ asymptotically by an
overall scale factor (Epistrophe Theorem, Paper I).

The minimal set of escape segments typically omits some segments
(strophes) that appear in the actual numerically-computed escape-time
plot.  Nevertheless, the results of Paper I apply to such strophic
segments as well.  There will be an epistrophe which converges upon an
endpoint of a strophe (Epistrophe Theorem).  However, we cannot in
general predict at which iterate such an epistrophe will begin.  On
the other hand, the numerical evidence of Fig.~\ref{fETP} and of
Fig.~2 in Paper I is suggestive that even in this case, the
epistrophes often begin $\Delta$ iterates beyond the strophe.

Strophes occur due to structure in the lobes $E_{-n}$ that we have
ignored in our simple topological picture of the tangle.  For example,
$E_n$ may develop additional ``fingers'' or ``branches'' as it is
mapped backwards.  These fingers spread out into the phase space,
creating additional intersections with the line of initial conditions.
(In some cases, such fingers can be connected with the presence of an
island chain inside the complex, such as the prominent period-5 chain
in Fig.~\ref{fPSP}.)  In general, a countable infinity of topological
parameters are needed to completely describe the
fingers\cite{Easton86,Rom-Kedar90,Rom-Kedar94,Ruckerl94a}, though we
expect a finite number of parameters to suffice for the escape-time
plot up to a given finite number of iterates.  The homotopy formalism
presented here can be generalized to incorporate these additional
topological parameters, thereby predicting at least some of the
strophe segments.  We will address these issues in a future paper.

In future work, we will also study winding numbers, explaining the
patterns shown in Fig.~\ref{fETP}.  In addition, we will apply our
results to the ionization of hydrogen in parallel electric and
magnetic fields.

\section{Acknowledgments}

The authors would like to thank Prof. Nahum Zobin for many useful
discussions.  This work was financially supported by the National
Science Foundation.

\appendix

\section{Proof of Algorithm~\ref{a1}}

\label{sAlgProof}

We need only verify statements A and B in Algorithm~\ref{a1}.  These
are certainly true when all elements of the untangled basis are
allowed in the expansion of $\ell_N$ (i.e., we do not omit the factors
specified in Step 1.)  This fact is evident by simply considering how
a path is constructed from a reduced product of the basis paths shown
in Fig.~\ref{fBC}; at each occurence of a $u_n$-factor, the path must
cross under the hole and hence through $E_n$, thus creating an escape
segment at the specified location (and with the specified direction.)
The $u_n$-factors are thus the key to determining the minimal set of
escape segments.  The $c_n$ basis elements ($n \le D$) create new
$u_n$-factors via Eqs.~(\ref{r8}) and (\ref{r15}) and are thus
themselves critical in determining the minimal set.  However, the
$e_n$ ($n \ge 0$) and $s_n$ ($-\infty < n < \infty$) basis elements
are ``inert'', mapping forward via Eqs.~(\ref{r11}) and (\ref{r12}),
never producing any $u_n$-factor.  We thus lose nothing by eliminating
them altogether from any expression which might contain them, as we
have done in Eqs.(\ref{r2}) -- (\ref{r3}).  (One can verify that
making these eliminations does not produce spurious cancellations of
$c_n$- or $u_n$-factors.)

\section{Proof of Theorem~\ref{t1}} 

\label{sT1proof}

Defining the two path-classes 
\begin{eqnarray}
\gamma & = & u_0 c_1 u_1, 
\label{r20} \\
\eta & = & u_0^{-1} F u_\Delta^{-1} F^{-1} u_0,
\label{r21}
\end{eqnarray}
we have the following lemma (recalling that all $s_n$ and $e_n$ basis
elements are omitted from our formulas.)

\noindent \textit{Lemma: For any $N \ge N_0$, $\ell_N$ can be
expressed as a product of elements in the set $S = ( c_1, ..., c_D \;
; \; \Box, \Box, u_2, ..., u_D, \Box, u_{D+2}, ... \; ; \; \gamma, \eta )$,
assuming $D>1$; for $D=1$, $S = ( c_1 \; ; \; \Box, \Box, \Box, u_{3},
u_4, ... \; ; \; \gamma, \eta )$.  (The symbol $\Box$ emphasizes the absence
of the classes $u_0$, $u_1$, $u_\Delta$.)}

\noindent \textit{Proof of Lemma:} It follows from the propagation
formulas (\ref{r8}), (\ref{r6}), and (\ref{r2}) and the definitions
of $n_c$ and $n_u$ that for $N \ge \max \{ -n_c+1, -n_u, 0 \}$,
$\ell_N$ can be expressed as a product of the elements $( c_1, ...,
c_D \; ; \; u_0, u_{1}, ...  )$.  Thus, for $N \ge N_0$, $\ell_N$ can
be expressed as a product of the elements $( c_{D+3}, ..., c_{2D+2} \;
; \; u_{D+2}, u_{D+3}, ...  )$.  Since the elements $u_{D+2}, u_{D+3},
...$ are in the set $S$, we need only verify that the elements
$c_{D+3}, ..., c_{2D+2}$ can be expressed as products of elements in
$S$, a fact which follows from rewriting Eqs.~(\ref{r22}) --
(\ref{r23}) as
\begin{subequations}
\begin{eqnarray}
c_{D+3} & = & (F^{-1} \gamma) (c_2 u_2^{-1} c_2^{-1}) (\gamma^{-1} F) , \\
c_{D+4} & = & (F^{-1} \gamma c_2 u_2) 
(c_3 u_3^{-1} c_3^{-1}) 
(u_2^{-1} c_2^{-1} \gamma^{-1} F), \\
& \vdots & \nonumber \\
c_{D+n} & = & (F^{-1} \gamma c_2 u_2 ... c_{n-2} u_{n-2}) 
(c_{n-1} u_{n-1}^{-1} c_{n-1}^{-1}) 
(u_{n-2}^{-1} c_{n-2}^{-1} ... u_2^{-1} c_2^{-1} \gamma^{-1} F), \\
& \vdots & \nonumber \\
c_{2D+1} & = & (F^{-1} \gamma c_2 u_2 ... c_{D-1} u_{D-1}) 
(c_D u_D^{-1} c_D^{-1}) 
(u_{D-1}^{-1} c_{D-1}^{-1} ... u_2^{-1} c_2^{-1} \gamma^{-1} F), \\
c_{2D+2} & = & (F^{-1} \gamma c_2 u_2 ... c_D u_D )
(F^{-1} \eta F)
( u_D^{-1} c_D^{-1} ... u_2^{-1} c_2^{-1} \gamma^{-1} F), 
\end{eqnarray}
\end{subequations}
for the case $D>1$.  For $D=1$, Eq.~(\ref{r34}) yields $c_4 = c_1^{-1}
\gamma c_1^{-1} \eta c_1 \gamma^{-1} c_1$.  This completes the proof of the lemma.

\vskip 12pt

For $N \ge N_0$, we expand $\ell_N$ as a product of elements in $S$.
By using Eqs.~(\ref{r20}) and (\ref{r21}) to eliminate $\gamma$ and $\eta$, we
obtain the expansion of $\ell_N$ in the untangled basis.  It is easy
to verify that when making these substitutions, there are no
cancellations of any $u_n$-factors.  (Here, we use the fact that
powers of $\eta$, such as $\eta^2$, can not occur in the expansion of
$\ell_N$ since this would imply that $\mathcal{L}_0$ has a
self-intersection.)

The theorem is now a trivial consequence of the representation of
$\ell_N$ as a product of elements in the set $S$.  Specifically, each
occurrence of $\gamma$ in the product yields a single segment which escapes
at $N-1$ iterates, corresponding to the $u_1$-factor of $\gamma$ in
Eq.~(\ref{r20}), and a single segment which escapes at $N$ iterates,
corresponding to the $u_0$-factor of $\gamma$.  The form of Eq.~(\ref{r20})
also implies that the directions and relative positions of these two
segments obey the Epistrophe Continuation Rule.

Similarly, Eq.~(\ref{r21}) implies that each occurrence of $\eta$ in the
representation of $\ell_N$ yields a segment which escapes at
$N-\Delta$ iterates and two segments which escape at $N$ iterates; the
directions and relative positions of these segments obey the
Epistrophe Start Rule.  Since the basis elements $u_0$, $u_1$, and
$u_\Delta$ occur in the expansion of $\ell_N$ only within the $\gamma$- and
$\eta$-factors, these two rules completely determine the minimal set of
escape segments at $N$ iterates. $\mathcal{QED}$

\begin{figure}
\caption{\label{fPSP} A phase space portrait is shown for our 
saddle-center map, which possesses a homoclinic tangle.  The line of
initial conditions $\mathcal{L}_0$ coincides with $q = 1.72$.  }
\end{figure}

\begin{figure}
\caption{\label{fETP} Escape data $n_i$ and $n_w$ are plotted for the
saddle-center map in Fig.~\ref{fPSP}.  On the right, the number of
iterates $n_i$ required to escape is plotted as a function of $p$
parameterizing the line of initial conditions $\mathcal{L}_0$.  The
escape segment marked by an asterisk at $n_i = 15$ is the first
numerically computed segment that is not in the minimal set; it is a
strophe.  On the left is plotted the winding number of the trajectory
as it escapes to infinity.}
\end{figure}

\begin{figure}
\caption{\label{fSOSD1} Qualitative phase space portrait for the delay
time $D=1$.  Capture zone $C_1$ overlaps escape zone $E_{-1}$, so some
orbits enter the complex on one iterate and leave on the next.  The
complex is bounded by the unstable manifold $\mathcal{U}$ from
$\mathbf{z}_\textsf{X}$ to $\mathbf{P}_0$ and by the stable manifold
$\mathcal{S}$ from $\mathbf{P}_0$ back to $\mathbf{z}_\textsf{X}$.
The active region $A$ is the union of the complex with $E_n$, $n \ge
0$, and $C_n$, $n \le 0$.  The boundary $\partial A$ of $A$ contains
alternating segments $\mathcal{S}_n$ and $\mathcal{U}_n$ of the stable
and unstable manifolds.  The inner boundaries of $E_n$ and $C_n$ are
respectively denoted $\mathcal{E}_{n}$ and $\mathcal{C}_{n}$.  The
intersection of $C_n$ with $E_{n-2}$ is the ``hole'' $H_{n-2}$.}
\end{figure}

\begin{figure}
\caption{\label{fSOSD3} Qualitative phase space portrait for delay
time $D=3$.  Here $C_2$ overlaps $E_{-2}$.  $\mathcal{C}_n$ and
$\mathcal{S}_n$ link $\mathbf{Q}_{n-1}$ to $\mathbf{P}_n$ encircling
hole $H_{n-4}$; likewise, $\mathcal{U}_n$ and $\mathcal{E}_n$ link
$\mathbf{P}_n$ to $\mathbf{Q_n}$ encircling hole $H_n$.}
\end{figure}

\begin{figure}
\caption{\label{fBC} Basis paths are shown for the active region $A^*$
with an infinite number of holes $H_n$ punctured in it.  $A^*$ is
bounded below by $\partial A$, which has been straightened into a
line.  The basis of $\Pi(A^*, \alpha)$ contains the path-classes $s_n$
and $u_n$ ($-\infty < n < \infty$) which link $\mathbf{Q}_{n-1}$ to
$\mathbf{P}_n$ and $\mathbf{P}_n$ to $\mathbf{Q}_n$ along $\partial
A$.  The basis also includes the path-classes $c_n$, $-\infty < n \le
D$, (bounding capture zones) and $e_n$, $0 \le n < \infty$ (bounding
escape zones).  The classes $e_n$, $0 \le n$, encircle the holes
$H_n$, and the classes $c_n$, $n \le D$, encircle the holes $H_{-D+n
-1}$.  (Note that the ordering of the holes shown in the diagram does
not coincide with the order of their indices.)  For $D=1$ or $3$, the
reader may verify that the curves drawn above agree topologically with
those in Figs.~\ref{fSOSD1} and \ref{fSOSD3}.}
\end{figure}

\begin{figure}
\caption{\label{fdir} We illustrate the convention for assigning a
direction to each escape segment.  Each of the two bold segments shown
has a 1st and 2nd endpoint; the 1st endpoint precedes the 2nd endpoint
as one moves forward along the path $\mathcal{E}_n'$.  The direction
of each segment points from the 2nd to the 1st endpoint.}
\end{figure}

\begin{figure}
\caption{\label{fETPQ} The escape-time plots are shown qualitatively
for $\ell_0 = c_D$ where (a) $D = 1$ and (b) $D$ is arbitrary.  At
each $n_i$, a segment (in the minimal set) that escapes in $n_i$
iterates is represented by an arrow giving the direction of the
segment; $\ell_0$ itself points up.  The minimal set is determined by
Eq.~(\ref{r34}) for (a) and Eq.~(\ref{r23}) for (b).  Note that two
new segments are spawned $\Delta = D+1$ iterates beyond the first
segment, an example of the Epistrophe Start Rule. }
\end{figure}

\begin{figure}
\caption{\label{fSETP} The escape-time plots are shown qualitatively
for two example lines of initial conditions.  Figures (a) and (b) are
determined by Eqs.~(\ref{r50}) and (\ref{r54}), respectively.  In each
plot we indicate the value of $N_0$ after which all segments can be
deduced using Theorem \ref{t1}.}
\end{figure}

\begin{table}
\caption{\label{table1} Notation Summary}
\begin{ruledtabular}
\begin{tabular}{ll}
$\mathcal{M}$ & dynamical map \\
$\mathbf{z}_\textsf{X}$ & unstable fixed point \\
$\mathcal{S}$, $\mathcal{U}$ & tangled branches of the stable and unstable manifolds \\
$\mathbf{P}_n$, $\mathbf{Q}_n$ $(-\infty < n < \infty)$ 
& homoclinic intersections \\
$\alpha$ & set of all $\mathbf{P}_n$'s and $\mathbf{Q}_n$'s \\
$E_n$, $C_n$ $(-\infty < n < \infty)$ & escape and capture zones \\
$\mathcal{L}_0$ & line of initial conditions \\
$\ell_0$ & path-class of $\mathcal{L}_0$ \\
$D$ & minimum delay time of the complex \\
$A$, $\partial A$ & active region and boundary of the active region \\
$H_n$ $(-\infty < n < \infty)$ & holes in the active region \\
$A^* = A\setminus \cup_n H_n$ & active region minus holes \\
$\mathcal{E}_n$, $\mathcal{U}_n$ 
& paths along $\mathcal{S}$- and $\mathcal{U}$-boundaries of $E_n$ \\
$\mathcal{S}_n$, $\mathcal{C}_n$ 
& paths along $\mathcal{S}$- and $\mathcal{U}$-boundaries of $C_n$ \\
$e_n, u_n, s_n, c_n$ & path-classes of $\mathcal{E}_n$, $\mathcal{U}_n$, $\mathcal{S}_n$, $\mathcal{C}_n$ \\
$F$ & path-class $c_1 c_2 ... c_D$ \\
$\Pi(A^*, \alpha)$ & fundamental groupoid of path-classes in $A^*$ having base points in $\alpha$ 
\end{tabular}
\end{ruledtabular}
\end{table}

\printfigures


\begin{thebibliography}{99}
\bibitem{footnote10} In this paper, we do not require
area-preservation and analyticity of $\mathcal{M}$, as in Assumption 1
of Paper I, but only that $\mathcal{M}$ be $C^1$.
\bibitem{footnote1} Topologically, the boundary of $A$ would be
defined as its closure minus its interior and would therefore include
the untangled branches of the stable and unstable manifolds.  Here,
when we say the boundary of $A$ we mean only the outer boundaries of
the capture and escape zones as well as the \textsf{X}-point.
\bibitem{footnote2} More formally, two (continuous) paths $\mathcal{A}_1,
\mathcal{A}_2: I \rightarrow A^*$ (where $I = [0,1]$) are homotopic if
there exists a continuous function $H:I\times I \rightarrow A^*$ such
that $\forall t \in I$, $H(0,t) = \mathcal{A}_1(t)$, $H(1,t) =
\mathcal{A}_2(t)$ and $\forall \tau \in I$, $H(\tau,0) = \mathbf{s}_0 =
\mathcal{A}_1(0) = \mathcal{A}_2(0)$, $H(\tau,1) = \mathbf{s}_1 =
\mathcal{A}_1(1) = \mathcal{A}_2(1)$.  The map $H$ is called a
homotopy.
\bibitem{footnote7} In practice, the line of initial conditions
typically extends beyond the active region; here, we truncate it and
consider only the piece inside $A$.  In practice, the line of initial
conditions may also pass through a hole $H_n$ or ${\mathbf
z}_{\textsf{X}}$; here we assume for simplicity that it does not.  In
Sect.~\ref{sExample2} we consider an example in which $\mathcal{L}_0$
does intersect a hole $H_n$.
\bibitem{footnote3} Each endpoint can be shifted either forward or
backward along $\partial A$.  If an endpoint lies on $\mathcal{S}_n$,
for some $n$, this choice does not affect the minimal set of escape
segments.  However, if an endpoint lies on $\mathcal{U}_{n}$, then
that end of $\mathcal{L}_0$ terminates inside $E_n$, meaning that
$\mathcal{L}_0$ ends in an escape segment.  Depending upon the
direction in which we shift this endpoint, we can either include or
exclude this terminal segment from the minimal set.  By convention, we
shift the endpoint in the direction which includes the terminal
segment.  See the examples in Sect.~\ref{sExamples}.
\bibitem{footnote8} One may ask whether it is even possible to choose
curves $\mathcal{L}_0'$, $\mathcal{E}_{-n_1}'$, and
$\mathcal{E}_{-n_2}'$ satisfying the conditions of Question 2?  That
this is indeed possible is established by the following lemma, which
is a simple corollary to a theorem of Turaev (Theorem 2 of
Ref.~\cite{Turaev79}.)
%
\begin{lemma}
Given a set of homotopy classes $a_i \in \Pi(A^*, \alpha)$, $i = 1,
..., m$, there exists a choice of representatives $\mathcal{A}_i \in
a_i$, $i = 1, ..., m$, such that for each $i$, the number of
self-intersections of $\mathcal{A}_i$ is minimized and for each pair
$i,j$, the number of pairwise-intersections between $\mathcal{A}_i$
and $\mathcal{A}_j$ is minimized.
\end{lemma}
\bibitem{footnote9} More generally, for a path $\mathcal{A}$ that does
not intersect $\mathbf{z}_{\mathsf{X}}$ and that has a well-defined
homotopy class in $\Pi(A^*, \alpha)$, the homotopy class of
$\mathcal{A}$ has a (unique) finite reduced expansion in the basis.
\end{thebibliography}
\end{document}